\documentclass[journal]{IEEEtran}
\usepackage{float}
\usepackage{amsmath,amssymb,amsfonts,mathtools}
\usepackage{algorithm}
\usepackage{algpseudocode}
\usepackage{graphicx}
\usepackage{array}
\usepackage{stfloats}
\usepackage{cite}
\usepackage{url}
\usepackage{textcomp}
\usepackage{booktabs}
\usepackage{tabularx}
\usepackage{multirow}
\usepackage{xcolor}
\usepackage{placeins}

\definecolor{wheat}{rgb}{0.96,0.87,0.70}

\definecolor{mario}{rgb}{0.8,0.8,1}

\definecolor{filippo}{rgb}{1,0.8,0.1}

\begin{document}
	\bstctlcite{IEEEexample:BSTcontrol}
	\title{Dynamic Coordination of Autonomous Vehicles at Intersections}
	\author{Saugat Shahi, Filippo Fabiani and Mario Zanon}
	
	\maketitle

\begin{abstract}
The coordination of autonomous vehicles at unsignalized intersections requires deciding both the trajectories of the vehicles and their crossing order, with the additional difficulty that the set of vehicles to be coordinated changes as traffic flows through the intersection. Since the resulting mixed-integer optimal control problem is hard to solve online, we propose an arrival-triggered receding-horizon scheme built around a schedule-refinement heuristic: for a fixed crossing order, a single nonlinear program (NLP) optimizes all trajectories and crossing times, and the first-order sensitivities of the cost with respect to the vehicle schedule are used to select adjacent swaps that are likely to reduce the overall cost. Replanning is triggered only when a new vehicle arrives, the new vehicle is appended at the tail of the current crossing order, and a memory of evaluated swaps that persists across replanning steps avoids redundant computations. Simulations of traffic scenarios involving streams of a few hundred heterogeneous vehicles show that the proposed scheme recovers most of the optimality gap incurred by first-in-first-out scheduling.
\end{abstract}

\section{Introduction}
Unsignalized intersections are among the traffic settings in which autonomous vehicles promise the largest benefits: by negotiating the crossing directly, without traditional traffic control, with substantial gains in travel time, energy consumption, and throughput~\cite{fabiani2019mixed, levin2017conflict, hult2019optimal}, which requires solving a problem that couples three decisions: a continuous trajectory planning, discrete crossing order scheduling, and since traffic evolves continuously revising both whenever a vehicle enters or leaves the intersection area.

Available work provides strategies at both ends of the complexity spectrum, see, e.g., the surveys~\cite{namazi2019intelligent, wu2022intersection,
rios2016survey, al2022signalized}. At one end, mixed-integer
formulations~\cite{fayazi2018mixed, muller2016intersection, fayazi2017optimal,yu2018integrated} treat the crossing order as binary variables optimized jointly with the trajectories whose exact solution grows combinatorially with the number of vehicles,increasing the complexity. At the other end, rule-based policies~\cite{qian2014priority, zhang2015state} and reservation-based schemes~\cite{dresner2008multiagent, levin2016optimizing} determine the order based on fixed criteria—predefined priorities or first-come-first-served slot allocation and therefore accommodate arrivals effortlessly. Being blind to the vehicle dynamics and to the individual objectives, however, these strategies can yield largely suboptimal schedules whenever the fleet is heterogeneous, as discussed in~\cite{shahi2026lagrange}.

Similar to~\cite{Hult2016PrimalDecomp, Zanon2017Asynchronous, hult2019optimal,
Hult2018ECC, Hult2020Coordination, Hult2020Experimental,
Hult2022Intersection}, we cast coordination as an optimal control problem in which collisions are excluded through precedence constraints on the intersection entry and exit times of consecutive vehicles. Related formulations~\cite{Murgovski2015Convex,deCampos2014Cooperative, KimKumar2014,Katriniok2017Distributed}, energy-optimal variants~\cite{zhang2016optimal,malikopoulos2018decentralized, rios2017online}, and multi-intersection extensions~\cite{chalaki2021optimal} follow a similar principle, though they consider the fixed crossing order. Schedule-refinement heuristics improve on this by updating the order based on information returned by the motion
planner via, e.g., mixed-integer quadratic approximation~\cite{Hult2018CDC}, dynamically reassigned priorities within prioritized distributed model predictive control (MPC)~\cite{Scheffe2022DynamicPriority}, or online reordering in mixed traffic~\cite{Faris2024TCST}. In~\cite{shahi2026lagrange}, we followed this route using duality:
parametric sensitivities are exploited to select the pair of successive vehicles in the schedule for which relaxing the precedence constraint would yield the largest gain in terms of cost; this information is used to guide attempts at swapping the order of such pairs with the aim of reducing the cost.

A possible framework for a dynamic setting is receding-horizon MPC, which introduces feedback through repeated optimization and thereby adapts to evolving traffic~\cite{funke2017, zhou2019distributed, Katriniok2017Distributed}. Yet, in the available MPC-based coordination schemes, the crossing order is either frozen along the horizon or provided by an external scheduler that is independent to the trajectory optimization; whenever the order changes, the controller must be re-initialized, and part of the achievable performance is lost.

In this paper, we bridge schedule refinement and feedback by
embedding the multiplier-guided reordering of~\cite{shahi2026lagrange} into an arrival-triggered receding-horizon scheme. Whenever a vehicle enters the
coordination area, it is appended at the tail of the current crossing order, a single NLP re-optimizes the trajectories and crossing times of all active vehicles, and the precedence multipliers rank the candidate adjacent swaps. Because the multipliers only carry local information, a swap is committed only if it verifiably reduces the cost. Two mechanisms keep the effort per
replanning event bounded: each NLP is warm-started from the previous primal-dual solution, in the spirit of real-time
iteration scheme~\cite{diehl2005real}, and a memory of already-evaluated swap pairs is carried across replanning steps, so that already-evaluated pairs are not re-tested. Simulations spanning randomized six- and nine-vehicle scenarios and sustained streams of several hundred heterogeneous
vehicles show that the proposed scheme recovers, on average, more than $85\%$ of the optimality gap incurred by first-in-first-out scheduling, and that this advantage is preserved as congestion increases, with a number of
NLP solves per arrival that remains small, making the
approach well-suited to online unsignalized intersection coordination. The computationally efficient implementation of our algorithm is beyond the scope of this paper and the subject of ongoing research.

\subsubsection*{Notation}
The set of consecutive integers from $a$ to $b$ is denoted
$\mathbb{I}_a^b \coloneqq \{a, a+1, \ldots, b\}$, with $\mathbb{I}_a^b = \emptyset$ if $a > b$.
For a finite set $\mathcal{S}$, $|\mathcal{S}|$ denotes its cardinality. Throughout the paper, $k$ denotes the discrete time step, $n$ the prediction step within the horizon, $t_\mathrm{s}$ the sampling time, $N$ the prediction horizon, and $M$ the maximum number of vehicles; the superscript $\dagger \in \{\mathrm{in}, \mathrm{out}\}$ is a placeholder for entry and exit quantities,
so that any statement involving $\dagger$ holds for both values.  For each vehicle $i$, the state and input trajectories are collected as $x_i \coloneqq (x_{i,0}, \ldots, x_{i,N})$ and $u_i \coloneqq (u_{i,0}, \ldots, u_{i,N-1})$, indexed by the prediction step $n$.
A \emph{schedule} (crossing order) $S$ is a permutation of the vehicles to be coordinated, $S_c$ being the vehicle in the $c$-th position. Superscript $(r)$ denotes quantities at iteration $r$ of the schedule-refinement procedure,
while $\cdot^\star$ denotes optimal quantities.

\section{Problem formulation}

\begin{figure}[t]
    \centering
    \includegraphics[width=1\linewidth]{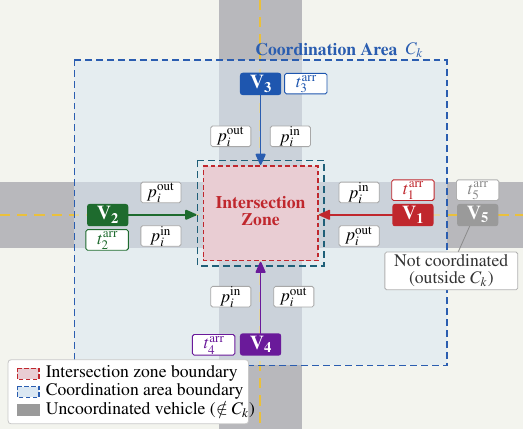}
    \caption{Four-way unsignalized intersection scenario. Vehicles $V_1$--$V_4$ 
    arrive asynchronously from different directions at times 
    $t_1^{\mathrm{arr}},\ldots,t_4^{\mathrm{arr}}$ and travel along fixed 
    straight paths toward the intersection zone (dashed red boundary). Entry 
    and exit positions are denoted by $p_i^\mathrm{in}$ and $p_i^\mathrm{out}$, 
    respectively, and colored arrows indicate each vehicle's direction of travel. Vehicle $V_5$ has not yet entered the coordination area $\mathcal{C}_k$ (dashed blue boundary) and is therefore not coordinated.}
    \label{fig:intersection_diagram}
\end{figure}

In this paper, we aim at solving the problem of optimally coordinating  cooperative autonomous vehicles approaching an unsignalized intersection, as illustrated 
in Figure~\ref{fig:intersection_diagram}. While we assume that each vehicle follows a predefined path without changing lane, we will consider the case in which the set of vehicles inside the coordination area changes over time, thus including possible new arrivals and departures.

Each vehicle $i$ has state $x_{i,k}\in\mathbb{R}^{n_{x_i}}$ and control input 
$u_{i,k}\in\mathbb{R}^{n_{u_i}}$ at time step $k$, with dynamics governed by:
\begin{equation}
    x_{i,k+1} = f_i(x_{i,k}, u_{i,k}),
    \label{eq:dynamics_function}
\end{equation}
which we assume to be twice continuously differentiable, and shall satisfy predefined path constraints:
\begin{equation}
    h_i(x_{i,k}, u_{i,k}) \le 0.
    \label{eq:path_constraints}
\end{equation}
Each vehicle enters and exits the intersection at fixed positions 
$p_i^{\mathrm{in}}$ and $p_i^{\mathrm{out}}$, with entry and exit times $t_i^{\mathrm{in}}$ and $t_i^{\mathrm{out}}$ 
defined implicitly via
\begin{equation}
p_i(t_i^{\dagger},x_i,u_i) = p_i^{\dagger}, \qquad 
    \dagger \in \{\mathrm{in},\mathrm{out}\},
    \label{eq:entry_exit_times}
\end{equation}
where $p_i(t, x_i, u_i)$ denotes the position of vehicle $i$ at time $t$, 
obtained through~\eqref{eq:dynamics_function} along 
the trajectory $(x_i, u_i)$.
Collision-free operation is then enforced via precedence constraints
\begin{equation}
    t_j^{\mathrm{in}} \geq t_i^{\mathrm{out}},
    \label{eq:precedence_pairwise}
\end{equation}
imposing that vehicle $i$ is scheduled to cross before vehicle $j$.

Let $t_i^{\mathrm{arr}} \in \mathbb{R}_{\geq 0}$, $t_i^{\mathrm{exit}} \in \mathbb{R}_{\geq 0}$ denote respectively the \emph{arrival time} and  \emph{exit time} 
of vehicle $i$, i.e., the time at which it enters/exits the coordination area. 
At each time step $k$, with sampling time $t_\mathrm{s} > 0$, the \emph{coordination set} reads
\begin{equation}
    \mathcal{C}_k := \bigl\{ i : t_i^{\mathrm{arr}} \leq k t_\mathrm{s} \leq t_i^{\mathrm{exit}} \bigr\},
    \label{eq:coordination_set}
\end{equation}
which collects all vehicles that have arrived inside the coordination area but have not yet left it. 
The crossing order at time step $k$ is a permutation $S_k$ of the indices in $\mathcal{C}_k$, defining the sequence in which vehicles are scheduled to cross the intersection.

The optimization variables are the states $x$, controls $u$, and entry/exit times $t$ of all vehicles, defined as
\begin{align*}
    x &:= (x_i)_{i \in \mathcal{C}_k}, &
    u &:= (u_i)_{i \in \mathcal{C}_k}, &
    t &:= (t_i)_{i \in \mathcal{C}_k},
\end{align*}
where $t_i := (t_i^{\mathrm{in}},t_i^{\mathrm{out}})$. Given a coordination set $\mathcal{C}_k$ and order $S_k$, the coordination 
problem over the prediction horizon $n\in\mathbb{I}_0^N$ is formulated as:
\begin{subequations}
\label{eq:coordination_ocp}
\begin{alignat}{2}
    V(S_k) := \min_{x,u,t} \quad 
    & J(x,u) \label{eq:objective_function} \\[1ex]
    \mathrm{s.t.} \quad
    & x_{i,0} = \hat{x}_{i,0},
    && \forall\, i \in \mathcal{C}_k, \label{eq:init_conditions} \\
    & x_{i,n+1} = f_i(x_{i,n}, u_{i,n}),
    && \forall\, i \in \mathcal{C}_k,\, n \in \mathbb{I}_0^{N-1}, 
    \label{eq:dynamics_compact} \\
    & h_i(x_{i,n}, u_{i,n}) \le 0,
    && \forall\, i \in \mathcal{C}_k,\, n \in \mathbb{I}_0^{N-1}, 
    \label{eq:path_constraints_2} \\
    & t_{S_c}^{\mathrm{in}} \ge t_{S_{c-1}}^{\mathrm{out}},
    && \forall\, c \in \mathbb{I}_2^{|\mathcal{C}_k|}, 
    \label{eq:precedenceconstraint} \\
    & p_i(t_i^{\mathrm{in}},x_i,u_i) = p_i^{\mathrm{in}},
    && \forall\, i \in \mathcal{C}_k, 
    \label{eq:entry_position} \\
    & p_i(t_i^{\mathrm{out}},x_i,u_i) = p_i^{\mathrm{out}},
    && \forall\, i \in \mathcal{C}_k. 
    \label{eq:exit_position}
\end{alignat}
\end{subequations}
The cost $J(x,u) := \sum_{i \in \mathcal{C}_k} J_i(x_i,u_i)$ is the sum of 
individual vehicle costs
\begin{align*}
J_i(x_i,u_i) := \sum_{n=0}^{N-1} \ell_i(x_{i,n}, u_{i,n}) 
    + \ell_{i,\mathrm{f}}(x_{i,N}),
\end{align*}
where $\ell_i$ and $\ell_{i,\mathrm{f}}$ are the stage and terminal costs, 
respectively. Notice that~\eqref{eq:coordination_ocp} is an NLP with continuous optimization 
variables $(x, u, t)$, solved for a \emph{fixed} schedule $S_k$, where $\hat{x}_{i,0}$ denotes the state of vehicle $i$ at time step $k$, used as initial condition of the prediction. 

For a fixed coordination set, \eqref{eq:coordination_ocp} coincides with the lower-level NLP of~\cite{shahi2026lagrange}, and the joint optimization over both the schedule $S_k$ and the continuous variables $(x, u, t)$ is the MINLP studied there. In~\cite{shahi2026lagrange}, we showed that mixed-integer formulations of this problem are orders of magnitude slower than the NLP, even when the schedule is fixed, and that exhaustive enumeration of the $|\mathcal{C}_k|!$ schedules is only viable for a handful of vehicles. We therefore adopt the same strategy as in~\cite{shahi2026lagrange}: the schedule $S_k^{(r)}$ is updated at each
iteration $r$, and the continuous NLP~\eqref{eq:coordination_ocp} is solved
for each given schedule $S_k^{(r)}$ to evaluate its cost.
Once these iterations converge, or the available computational time is over, we denote the final
iteration as $r^\star$, and the final schedule is $S_k=S_k^{(r^\star)}$. To address the time-varying nature of $\mathcal C_k$, which introduces a key difference with respect to the static  setting of~\cite{shahi2026lagrange}, we note that vehicles entering or leaving the coordination area introduce a specific structure in the problem that can be exploited.  Indeed, as a vehicle leaves the coordination area, the optimal schedule cannot change, while when a vehicle enters the coordination area, it is unlikely that the schedule changes dramatically.
Since consecutive coordination 
sets $\mathcal{C}_{k-1}$ and $\mathcal{C}_k$ differ by at most one 
vehicle, the schedule $S_{k-1}$ computed at the previous replanning instant provides 
the initial schedule for the current problem, once the new vehicle is appended in the last position. This enables an  incremental and efficient update of the coordination policy.

\section{Arrival-triggered MPC with Lagrange-multiplier-based reordering}

We start by briefly recalling the key idea of the approach in~\cite{shahi2026lagrange} before describing the crucial extension proposed here.

In the static setting, i.e., $|\mathcal C_k|=M$ for all $k$, each vehicle has a given initial state, and~\eqref{eq:coordination_ocp} is solved once to compute a crossing order together with the corresponding optimal trajectories. The key idea is that the Lagrange multipliers associated with the precedence constraints~\eqref{eq:precedenceconstraint} carry first-order sensitivity information: the multiplier $\lambda_c \ge 0$ quantifies the marginal cost of enforcing the collision-avoidance constraint between vehicles $S_c$ and $S_{c+1}$, so that a large $\lambda_c$ indicates that swapping pair $c$ is more likely to reduce the cost than pairs with smaller $\lambda_c$. This intuition leads to an algorithm that sequentially attempts adjacent swaps in the schedule until no improvement can be achieved. The result is a computationally efficient heuristic that avoids the complexity of genuine mixed-integer solvers, at the price of accepting some degree of suboptimality, and that consistently delivers substantial improvements over FIFO scheduling in simulation.

In practice, vehicles do not all arrive simultaneously: they enter the coordination area one by one, and the problem solved for the previous set of vehicles must be updated online at each new arrival. A naive approach would discard the current solution and solve~\eqref{eq:coordination_ocp} from scratch every time a new vehicle appears. This is wasteful, as the schedule computed for the previous set of vehicles is unlikely to change dramatically. We
thus propose an \emph{incremental} strategy built on two ingredients: initialization, and swap memory.

    \paragraph*{Initialization} When a new vehicle arrives, it is appended at the end of the current crossing order $S_{k-1}$. This initialization is natural, since a vehicle that has just entered the coordination area is typically farther from the intersection than all other vehicles, and is therefore last in the FIFO order; the multiplier-guided swaps then move it forward whenever this reduces the cost. The NLP is initialized with a simple guess built from the current vehicle states: each vehicle, including the newly arrived one, accelerates towards and then cruises at its reference velocity, and the entry and exit times are set sequentially according to the crossing order.

    \paragraph*{Swap memory} A set $\mathcal{T}$ of previously evaluated swap pairs is maintained to exclude them from the candidate set, reducing the number of NLP solves at each replanning step. As we detail below, the way $\mathcal{T}$ is managed across replanning defines two variants of the inner loop, denoted H1 and H2. They are the online counterparts of Heuristic~1 and Heuristic~2 in~\cite{shahi2026lagrange}: as there, H1 never re-evaluates a pair that has already been evaluated, whereas H2 allows pairs to be re-evaluated. The only difference lies in the event that clears the memory of H2, which is an accepted swap in the static setting of~\cite{shahi2026lagrange} and a new arrival here.
Before introducing the algorithms, let us introduce the notation they employ. The set of newly arrived vehicles at step $k$ is $\mathcal{C}_k^{\mathrm{new}} := \mathcal{C}_k \setminus\mathcal{C}_{k-1}$, while vehicles that have left the coordination area, i.e., those in $\mathcal{C}_{k-1} \setminus \mathcal{C}_k$, are simply removed from the crossing order, which does not change the relative order of the remaining vehicles; and $u_{i,0}^\star$ denotes the first element of the optimal control sequence of vehicle $i$, which is applied to the plant in MPC fashion. In the inner loop, $S^\star, Z^\star, J^\star$ denote, respectively, the current best schedule, primal-dual solution, and cost, and $\lambda^\star$ the vector of precedence multipliers contained in $Z^\star$. The subscript $\cdot_{\mathrm{sw}}$ denotes the
corresponding quantities for the candidate swap under evaluation. The operator $\mathrm{swap}(S, c)$ returns the schedule obtained by exchanging element $c$ and $c+1$ of $S$; and $\mathcal{P}$ is the set of admissible swap indices. The parameter $\alpha \in \{1,2\}$ selects the heuristic variant, as detailed next.
\begin{algorithm}[t]
\caption{Outer loop: arrival-triggered receding-horizon coordination}
\label{alg:outer_mpc}
\small
\begin{algorithmic}[1]
\State \textbf{Input:} $t_\mathrm{s}, N, \{t_i^{\mathrm{arr}}\},
        \{t_i^{\mathrm{exit}}\}, \{\hat{x}_{i,0}\}, \alpha$
\State \textbf{Init:} $\mathcal{C}_{-1} \gets \emptyset,\;
        S_{-1} \gets [\,],\; \mathcal{T} \gets \emptyset,\; Z^\star \gets \emptyset$
\For{$k = 0, 1, \ldots$}
    \State $\mathcal{C}_k \gets \{i : t_i^{\mathrm{arr}}
            \le k t_\mathrm{s} \le t_i^{\mathrm{exit}}\}$
    \State $S_k \gets S_{k-1}$ restricted to the vehicles in $\mathcal{C}_k$
    \State $\mathcal{C}_k^{\mathrm{new}} \gets
            \mathcal{C}_k \setminus \mathcal{C}_{k-1}$
    \If{$\mathcal{C}_k^{\mathrm{new}} = \emptyset$}
        \State apply $u_{i,0}^\star$ from $Z^\star$, advance $x_i$,
               $\forall\, i \in \mathcal{C}_k$
        \State \textbf{continue}
    \EndIf
    \State append $\mathcal{C}_k^{\mathrm{new}}$ to the tail of $S_k$
    \State $(S_k, Z^\star, \mathcal{T}) \gets
            \textsc{ReplanAndSwap}(\mathcal{C}_k, S_k, \{\hat{x}_{i,0}\}, \mathcal{T}, \alpha)$
    \State apply $u_{i,0}^\star$ from $Z^\star$, advance $x_i$,
           $\forall\, i \in \mathcal{C}_k$
\EndFor
\end{algorithmic}
\end{algorithm}
\begin{algorithm}[t]
\caption{Inner loop: \textsc{ReplanAndSwap}}
\label{alg:replan_swap}
\small
\begin{algorithmic}[1]
\State \textbf{Input:} $\mathcal{C}_k, S, \{\hat{x}_{i,0}\}, \mathcal{T}, \alpha$
\State \textbf{Output:} $S^\star, Z^\star, \mathcal{T}$
\If{$\alpha = 2$}
    \State $\mathcal{T} \gets \emptyset$
\EndIf
\State $S^\star \gets S$
\State Solve NLP~\eqref{eq:coordination_ocp} for $S^\star$,
        obtain $(J^\star, Z^\star, \lambda^\star)$
\While{\textbf{true}}
    \State $\mathcal{P} \gets \{c \in \mathbb{I}_1^{|\mathcal{C}_k|-1} : \{S^\star_c, S^\star_{c+1}\}
            \notin \mathcal{T}\}$
    \If{$\mathcal{P} = \emptyset$} \textbf{break} \EndIf
    \State $c^\star \gets \arg\max_{c \in \mathcal{P}} \lambda^\star_c$
    \State $S_{\mathrm{sw}} \gets \mathrm{swap}(S^\star, c^\star)$
    \State Solve NLP~\eqref{eq:coordination_ocp} for $S_{\mathrm{sw}}$,
            obtain $(J_{\mathrm{sw}}, Z_{\mathrm{sw}}, \lambda_{\mathrm{sw}})$
    \If{$J_{\mathrm{sw}} < J^\star$}
        \State $(S^\star, J^\star, Z^\star, \lambda^\star) \gets
                (S_{\mathrm{sw}}, J_{\mathrm{sw}},
                Z_{\mathrm{sw}}, \lambda_{\mathrm{sw}})$
        \If{$\alpha = 1$} \State $\mathcal{T} \gets \mathcal{T} \cup \{\{S^\star_{c^\star}, S^\star_{c^\star+1}\}\}$ \EndIf
    \Else
        \State $\mathcal{T} \gets \mathcal{T} \cup
                \{\{S^\star_{c^\star}, S^\star_{c^\star+1}\}\}$
    \EndIf
\EndWhile
\State \textbf{return} $(S^\star, Z^\star, \mathcal{T})$
\end{algorithmic}
\end{algorithm}

The proposed scheme is organized as a two-level loop. The \emph{outer loop} (Algorithm~\ref{alg:outer_mpc}) runs at every sampling time $k t_\mathrm{s}$ and is responsible for managing the coordination set $\mathcal{C}_k$, detecting new arrivals, and applying the first control from the most recent solution. Replanning is triggered exclusively by new arrivals, so the computational load is proportional to the rate of incoming vehicles rather than to the sampling frequency. When a new vehicle joins $\mathcal{C}_k$, it is appended to the tail of $S_{k-1}$, and the \emph{inner loop} (Algorithm~\ref{alg:replan_swap}) is invoked to refine the schedule and recompute the trajectories.

This scheme is slightly idealized. Because the NLP solution time is not negligible, the initial state passed to the inner loop must be the state that each vehicle will be in at some future time $t_\mathrm{alg}$, by which the algorithm must terminate. If $t_\mathrm{alg}$ is too far ahead, the algorithm risks becoming
ineffective, as some vehicle may already have reached the intersection by that time. A practical implementation thus commits to a short $t_\mathrm{alg}$ and accepts to only partially solve the problem. Once $t_\mathrm{alg}$ is reached, the outer loop is re-invoked with an updated $t_\mathrm{alg}$ and initial
states, even if no new vehicle has arrived.

The inner loop refines the crossing order using the KKT multipliers $\lambda$ of the precedence constraints~\eqref{eq:precedenceconstraint} to rank candidate
adjacent swaps. The multiplier $\lambda_c \ge 0$ measures the first-order sensitivity of the optimal cost with respect to constraint $c$, i.e., it quantifies how much one gains by relaxing constraint $c$~\cite{nocedal2006numerical}: a large value flags the pair $(S_c, S_{c+1})$ as likely reducing the cost. The admissible pair with the largest $\lambda_c$ is swapped and the cost re-evaluated. The swap is accepted if it reduces the overall cost, and rejected otherwise.

Following~\cite{shahi2026lagrange}, two variants of the procedure are obtained by varying the parameter $\alpha \in \{1, 2\}$. They differ in two aspects only: which evaluated pairs are stored in $\mathcal{T}$, and when $\mathcal{T}$ is cleared. In both variants, a rejected pair is stored in $\mathcal{T}$, so that it is not re-evaluated within the same replanning step. With H1 ($\alpha = 1$), accepted pairs are stored as well, and $\mathcal{T}$ is never cleared: every pair is evaluated at most once over the whole simulation, regardless of whether its swap was accepted or rejected. With H2 ($\alpha = 2$), only rejected pairs are stored, and $\mathcal{T}$ is cleared at every new arrival: a rejected pair can be re-evaluated after a new arrival, and an accepted swap can later be reverted. Consequently, at every replanning step the candidate set of H1 is a subset of that of H2, so that H1 typically performs fewer NLP solves than H2. A detailed comparison is reported in Section~\ref{sec:numerical-simulations}.

\section{Numerical simulations}
\label{sec:numerical-simulations}
\begin{figure*}[!t]
    \centering
    \begin{minipage}[t]{\columnwidth}
        \centering
        \includegraphics[width=\linewidth]{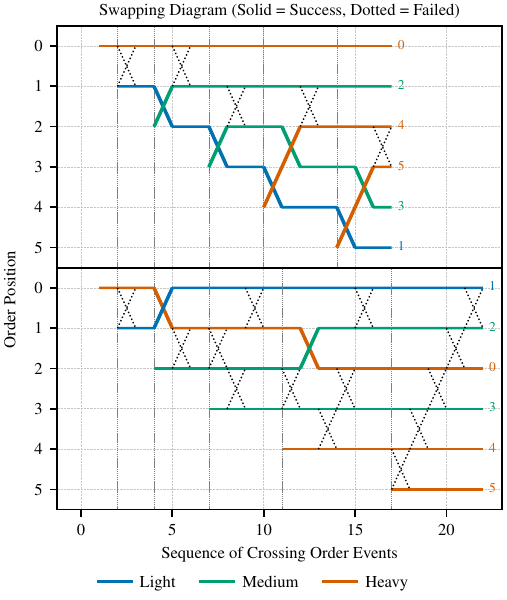}
        \caption{Scenario~1 (6 vehicles): swap diagrams under H1 (top) and H2 (bottom).}
        \label{fig:6s1}
    \end{minipage}\hfill
    \begin{minipage}[t]{\columnwidth}
        \centering
        \includegraphics[width=\linewidth]{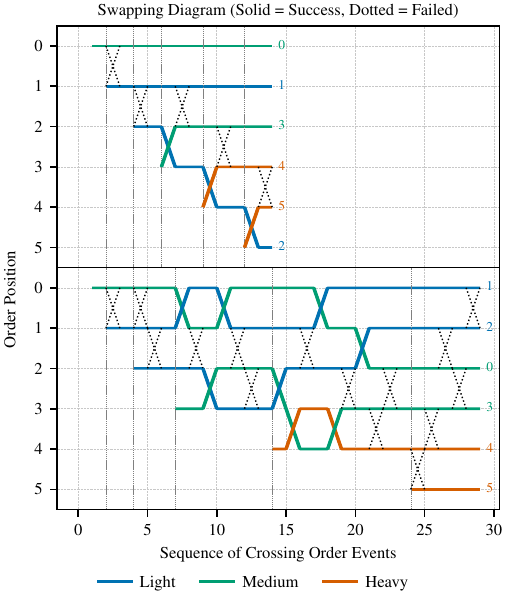}
        \caption{Scenario~2 (6 vehicles): swap diagrams under H1 (top) and H2 (bottom).}
        \label{fig:6s2}
    \end{minipage}
\end{figure*}
To assess the effectiveness of the proposed receding-horizon NLP framework 
with multiplier-guided swapping, we implemented a simulation environment in 
Python using CasADi~\cite{Andersson2019CasADi} and IPOPT~\cite{wachter2006implementation}. The setup considers an unsignalized intersection, where vehicles 
arrive at randomly generated times and the crossing order must be scheduled in real time.
The vehicle dynamics in~\eqref{eq:dynamics_function} are instantiated as a
discrete-time double integrator, so that the state evolves as:
\begin{equation}
\label{eq:dynamics_double_integrator}
\begin{aligned}
p_{i,k+1} &= p_{i,k} + t_\mathrm{s}\, v_{i,k}
            + \tfrac{1}{2}\, t_\mathrm{s}^{2}\, u_{i,k}, \\
v_{i,k+1} &= v_{i,k} + t_\mathrm{s}\, u_{i,k}.
\end{aligned}
\end{equation}

The initial positions and velocities are generated randomly,
while the entry and exit positions are fixed as \(p_i^{\mathrm{in}} = 0\) m and \(p_i^{\mathrm{out}} = 10\) m for all vehicles. 
The main parameters used throughout the experiments are summarized in Table~\ref{tab:system_parameters}.

\begin{table}[!tbh]
\centering
\caption{System Parameters}
\label{tab:system_parameters}
\begin{tabularx}{\columnwidth}{>{\hsize=0.6\hsize}X >{\hsize=1.6\hsize}X >{\hsize=0.8\hsize}X}
\toprule
\textbf{Parameter} & \textbf{Description} & \textbf{Value} \\
\midrule
$N$               & Prediction horizon              & 200 \\
$t_\mathrm{s}$        & Sampling time                              & 0.1 s \\
$M$               & Maximum number of vehicles                         & 6, 9 (up to 43 in Sec.~\ref{sec:zone_release}) \\
$p_i^{\mathrm{in}}$ & Entry position of vehicle $i$              & 0 m \\
$p_i^{\mathrm{out}}$& Exit position of vehicle $i$               & 10 m \\
$v_{\mathrm{ref}}$  & Reference velocity                         & 13.88 m/s \\
\bottomrule
\end{tabularx}
\end{table}

Each vehicle's cost function is quadratic and penalizes both the acceleration (control effort) and the deviation of the velocity from a reference value. In the considered simulation setting, the stage cost for vehicle $i$ is specifically chosen as
\begin{align}
    \ell_i(x_{i,n},u_{i,n}) = r_i\,u_{i,n}^2 + q_i\,(v_{i,n}-v_{\mathrm{ref}})^2,
\end{align}
while the terminal cost is defined as
\begin{align}
    \ell_{i,\mathrm{f}}(x_{i,N}) = q_i\,(v_{i,N}-v_{\mathrm{ref}})^2.
\end{align}

Larger values of \(r_i\) discourage aggressive control actions, while larger values of \(q_i\) enforce tighter tracking of the reference velocity. The coefficients \(q_i\) and \(r_i\) are selected according to three vehicle classes, summarized in Table~\ref{tab:vehicle_classes}, so as to introduce and evaluate a heterogeneous vehicle behavior within the same optimization framework.
\begin{table}[!tbh]
\centering
\caption{Vehicle classes and corresponding quadratic cost weights.}
\label{tab:vehicle_classes}
\begin{tabular*}{\columnwidth}{@{\extracolsep{\fill}}lcc}
\toprule
\textbf{Vehicle class} & \boldmath{$r_i$} & \boldmath{$q_i$} \\
\midrule
Light  & 1   & 1   \\
Medium & 10  & 10  \\
Heavy  & 100 & 500 \\
\bottomrule
\end{tabular*}
\end{table}

H1 and H2 are compared on the illustrative scenarios (Figs.~\ref{fig:6s1}--\ref{fig:9s2} and Tables~\ref{tab:S1_6Vehicle}--\ref{tab:S2_9Vehicle}) and on the randomly generated ones (Section~\ref{sec:random}), while the continuous-arrival experiment uses H1 only. In our simulations, the two heuristics do not always yield the same order and cost. H1 generally produces solutions of equal or lower cost than H2 with fewer NLP solves, since it does not attempt the swaps recorded in the swap memory~$\mathcal{T}$.
We will discuss next three sets of experiments. The first one includes simple scenarios that help us explain how the algorithm works. The second one considers random simulations that allow us to better assess the effectiveness of our algorithm on small-sized problems where we can use brute-force enumeration to obtain the optimal schedule. Finally, the third one considers a continuous stream of several hundred vehicles.

\subsection{Illustrative scenarios}
\begin{figure*}[!t]
    \centering
    \begin{minipage}[t]{\columnwidth}
        \centering
        \includegraphics[width=\linewidth]{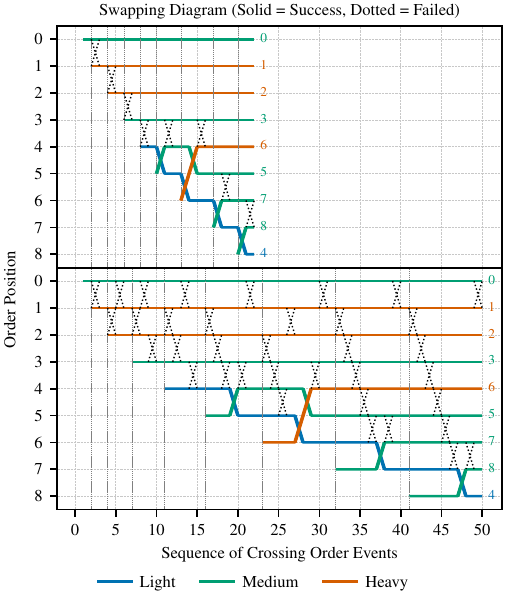}
        \caption{Scenario~1 (9 vehicles): swap diagrams under H1 (top) and H2 (bottom).}
        \label{fig:9s1}
    \end{minipage}\hfill
    \begin{minipage}[t]{\columnwidth}
        \centering
        \includegraphics[width=\linewidth]{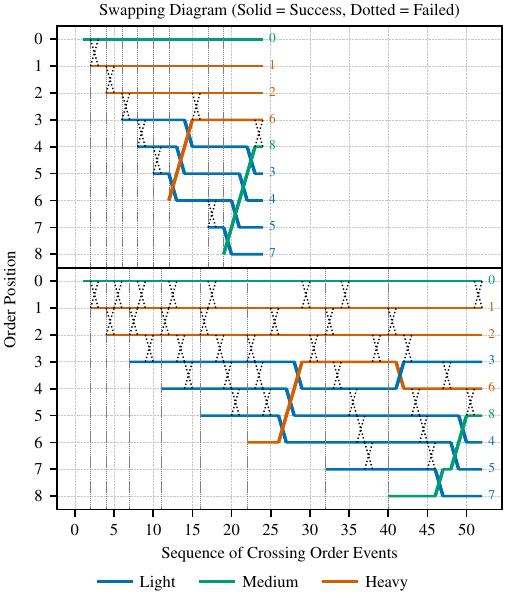}
        \caption{Scenario~2 (9 vehicles): swap diagrams under H1 (top) and H2 (bottom).}
        \label{fig:9s2}
    \end{minipage}
\end{figure*}
We demonstrate the behavior of Algorithm~\ref{alg:replan_swap} first on six-vehicle and then on nine-vehicle scenarios.We consider two six-vehicle scenarios with the same arrival times, $t^{\mathrm{arr}} = \{0.00, 0.36, 0.64, 0.99, 1.38, 1.69\}$\,s, but different fleet compositions, using the light, medium and heavy vehicle classes of Table~\ref{tab:vehicle_classes}, which differ in the cost weights $r_i$ and $q_i$. Vehicles are labeled $0,1,\ldots$ in order of arrival. In Scenario~1, the fleet is dominated by heavy vehicles: vehicles $0$, $4$ and $5$ are heavy, $2$ and $3$ medium, and $1$ light. In Scenario~2, the mix is more balanced: vehicles $4$ and $5$ are heavy, $0$ and $3$ medium, and $1$ and $2$ light. The class of each vehicle is also reported in the legend of the swap diagrams.

Figure~\ref{fig:6s1} shows the swap diagrams for
Scenario~1 under H1 (top) and H2 (bottom).
Each colored trace tracks a vehicle's position in the crossing order over the
sequence of replanning events triggered by Algorithm~\ref{alg:outer_mpc}. Solid lines display accepted swaps and dotted lines display rejected ones.
H2 generates visibly more attempted swaps, as previously evaluated pairs are re-evaluated after new arrivals, whereas H1 avoids those attempts by exploiting the memory stored in~$\mathcal{T}$. Further data is reported in Table~\ref{tab:S1_6Vehicle}.

Figure~\ref{fig:6s2} shows the swap diagrams
for Scenario~2.
Since vehicles $1$ and $2$, which arrive right after vehicle~$0$, are light, no swap is beneficial
until vehicle~$3$ (medium) arrives and is moved ahead of the light vehicle~$2$. Likewise, the arrival of each of the heavy vehicles $4$ and $5$ triggers one accepted swap, inserting the new vehicle ahead of the light vehicle~$2$. Further data is reported in Table~\ref{tab:S2_6Vehicle}.
\begin{table}[!tbh]
\centering
\caption{Analysis of Scenario-1 (6-Vehicle)}
\label{tab:S1_6Vehicle}
\begin{tabular*}{\columnwidth}{@{\extracolsep{\fill}}lcc}
\toprule
\textbf{} & H1 & H2 \\
\midrule
Cost & 16665 & 19741  \\
Number of swaps  & 11 & 16 \\
Crossing order  & [0, 2, 4, 5, 3, 1]   & [1, 2, 0, 3, 4, 5]   \\

\bottomrule
\end{tabular*}
\end{table}

\begin{table}[!tbh]
\centering
\caption{Analysis of Scenario-2 (6-Vehicle)}
\label{tab:S2_6Vehicle}
\begin{tabular*}{\columnwidth}{@{\extracolsep{\fill}}lcc}
\toprule
\textbf{} & H1 & H2 \\
\midrule
Cost & 9379 & 10614  \\
Number of swaps  & 8 & 23 \\
Crossing order  & [0, 1, 3, 4, 5, 2]   & [1, 2, 0, 3, 4, 5]   \\
\bottomrule
\end{tabular*}
\end{table}

We then consider two nine-vehicle scenarios with arrival times $t^{\mathrm{arr}} = \{0.00, 0.36, 0.64, 0.99, 1.38, 1.69, 1.99, 2.21, 2.46\}$\,s. In Scenario~1, vehicles $1$, $2$ and $6$ are heavy, $0$, $3$, $5$, $7$ and $8$ medium, and $4$ light; in Scenario~2, vehicles $1$, $2$ and $6$ are heavy, $0$ and $8$ medium, and $3$, $4$, $5$ and $7$ light.
Figures~\ref{fig:9s1} and~\ref{fig:9s2} show the swap diagrams for Scenario~1 and Scenario~2, respectively, with H1 at the top and H2 at the bottom of each figure. The dominant reordering event in both scenarios is the arrival of the heavy vehicle $6$, which triggers multiple accepted swaps and the largest single-step cost reduction in both cases.
Table~\ref{tab:S1_9Vehicle} and Table~\ref{tab:S2_9Vehicle} provide the detailed analysis of the two scenarios, respectively.

\begin{table}[!tbh]
\centering
\caption{Analysis of Scenario-1 (9-Vehicle)}
\label{tab:S1_9Vehicle}
\begin{tabular*}{\columnwidth}{@{\extracolsep{\fill}}lcc}
\toprule
\textbf{} & H1 & H2 \\
\midrule
Cost & 31895 & 31895  \\
Number of swaps  & 13 & 41 \\
Crossing order  & [0, 1, 2, 3, 6, 5, 7, 8, 4]   & [0, 1, 2, 3, 6, 5, 7, 8, 4]   \\
\bottomrule
\end{tabular*}
\end{table}

\begin{table}[!tbh]
\centering
\caption{Analysis of Scenario-2 (9-Vehicle)}
\label{tab:S2_9Vehicle}
\begin{tabular*}{\columnwidth}{@{\extracolsep{\fill}}lcc}
\toprule
\textbf{} & H1 & H2 \\
\midrule
Cost & 12435 & 13190  \\
Number of swaps  & 15 & 43 \\
Crossing order  & [0, 1, 2, 6, 8, 3, 4, 5, 7]   & [0, 1, 2, 3, 6, 8, 4, 5, 7]   \\
\bottomrule
\end{tabular*}
\end{table}

\subsection{Randomly-generated scenarios}
\label{sec:random}
To assess the proposed framework beyond individual scenarios, we conduct a statistical evaluation over two batches of 100 randomly generated six-vehicle simulation instances. In Batch~1, successive vehicle arrival gaps are drawn uniformly in the interval $[0.2, 0.4]$\,[s], producing dense traffic. In Batch~2, gaps are drawn uniformly in the interval $[0.3, 0.5]$\,[s], yielding more spread-out arrivals and less dense traffic than in Batch~1.
For each simulation instance, three strategies are compared: FIFO scheduling, the proposed H1/H2 heuristic, and brute-force enumeration of all $6! = 720$ permutations, which provides the global optimum as a reference. For each scenario and each strategy, the reported cost is the
optimal NLP value $V(S_k)$ of~\eqref{eq:coordination_ocp}
evaluated on the final crossing order produced by that strategy,
once all vehicles have arrived, taken at the terminal arrival. In this way, FIFO, H1/H2 heuristic, and brute-force enumeration are compared on the same metric.

\begin{figure}[!htb]
    \centering
    \includegraphics[width=\linewidth]{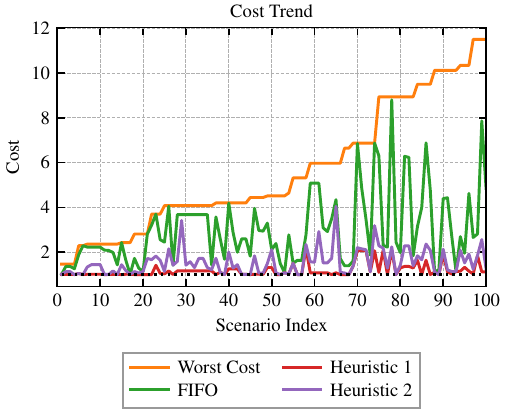}
    \caption{Cost trend across 100 scenarios (Batch~1, inter-arrival gaps
    $\in[0.2,0.4]$\,[s]). All costs are normalized by the brute-force (optimal) cost, shown as the black dotted line; Worst Cost is the largest cost over all $6!$ crossing orders, by which scenarios are sorted.
    }
    \label{fig:mpc1cost}
\end{figure}

Focusing on Batch~1, Figure~\ref{fig:mpc1cost} shows the cost across all 100 scenarios,
sorted by scenario worst cost. Our heuristics H1/H2 remain close to the optimal cost in most scenarios and well below the FIFO baseline. The worst-case cost is displayed to provide further information on how well or badly FIFO and H1/H2 perform: as opposed to H1/H2, FIFO can perform arbitrarily badly. As H1 yields lower or equal cost than H2 across all scenarios, H1 is used as the default heuristic for the evaluation of optimality gap reduction (OGR) and optimality gap (OG).

\begin{figure}[!htb]
    \centering
    \includegraphics[width=\columnwidth]{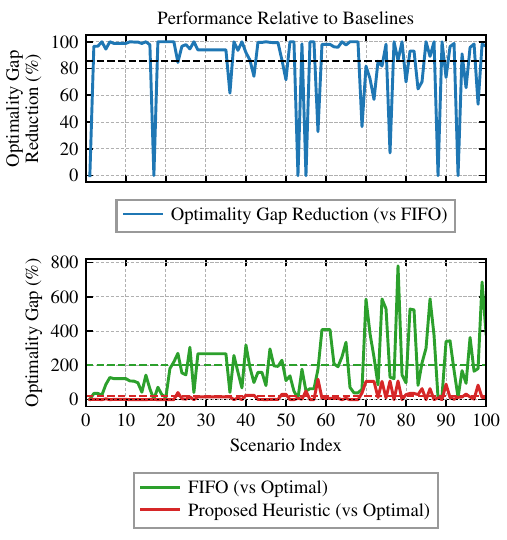}
    \caption{Performance metrics across 100 scenarios (Batch~1). Top:
    optimality gap reduction relative to FIFO (mean $85.5\%$). Bottom:
    optimality gap relative to brute-force optimal (mean $18.8\%$).}
    \label{fig:mpc1metrics}
\end{figure}

Let \(J_{\mathrm{BF}}^{(s)}\), \(J_{\mathrm{FIFO}}^{(s)}\), and \(J_{\mathrm{H}}^{(s)}\) denote the cost of brute force, FIFO and H1 respectively in scenario \(s\). The OGR is
\begin{equation}
\mathrm{OGR}^{(s)} =
\frac{J_{\mathrm{FIFO}}^{(s)} - J_{\mathrm{H}}^{(s)}}{J_{\mathrm{FIFO}}^{(s)} - J_{\mathrm{BF}}^{(s)}} \times 100,
\end{equation}
and the OG is
\begin{equation}
\mathrm{OG}^{(s)} =
\frac{J_{\mathrm{H}}^{(s)} - J_{\mathrm{BF}}^{(s)}}{J_{\mathrm{BF}}^{(s)}} \times 100.
\end{equation}
Figure~\ref{fig:mpc1metrics} reports the OGR relative to FIFO and the OG of H1.
The heuristic achieves a mean OGR of $85.5\%$, indicating that it recovers
the large majority of the gap between FIFO and optimal on average.
The mean OG of $18.8\%$ reflects the residual suboptimality accepted in
exchange for computational tractability; notably, several scenarios achieve
near-zero OG. 

\begin{figure}[!htb]
    \centering
        \includegraphics[width=\linewidth]{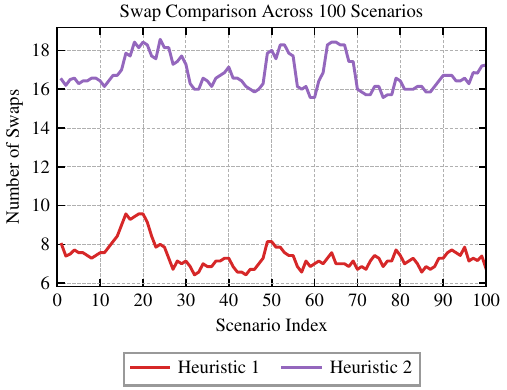}
        \caption{Swap count per scenario for H1 and H2 (Batch~1; moving average over 7 scenarios).
        }
        \label{fig:mpc1swaps}
\end{figure}

Figure~\ref{fig:mpc1swaps} compares H1 and H2 in terms of swap counts. H1 requires between 7 and 10 swap evaluations per scenario, whereas H2
performs between 15 and 19.

\begin{figure}[!htb]
    \centering
    \includegraphics[width=\linewidth]{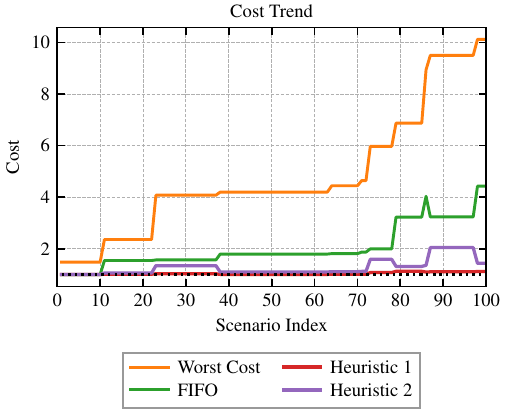}
    \caption{Cost trend across 100 scenarios (Batch~2, inter-arrival gaps
    $\in[0.3,0.5]$\,[s]). All costs are normalized by the brute-force (optimal) cost, shown as the black dotted line; Worst Cost is the largest cost over all $6!$ crossing orders, by which scenarios are sorted. 
    }
    \label{fig:mpc2cost}
\end{figure}
\begin{figure}[!htb]
    \centering
    \includegraphics[width=\linewidth]{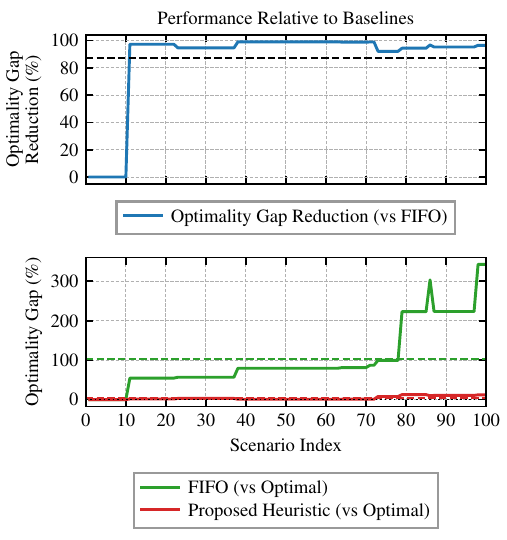}
    \caption{Performance metrics across 100 scenarios (Batch~2). Top:
    optimality gap reduction relative to FIFO (mean $87.0\%$). Bottom:
    optimality gap relative to brute-force optimal (mean $3.9\%$),
    substantially lower than Batch~1, reflecting reduced scheduling
    complexity under wider inter-arrival gaps.}
    \label{fig:mpc2metrics}
\end{figure}

By considering Batch~2, Figure~\ref{fig:mpc2cost} shows the cost across all $100$ scenarios.
The heuristic closely matches the optimal cost across all
instances and consistently performs better than FIFO, reproducing the qualitative pattern observed for Batch~1. Figure~\ref{fig:mpc2metrics} reports the OGR and OG for Batch~2.
The mean OGR of $87.0\%$ is marginally higher than in Batch~1,
while the mean OG drops substantially from $18.8\%$ to $3.9\%$.
This improvement reflects the effect of wider inter-arrival spacing, which makes the time-varying coordination problem simpler.

\begin{figure}[!htb]
    \centering

        \includegraphics[width=\linewidth]{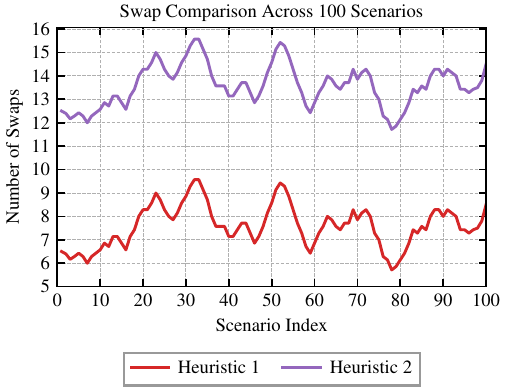}
        \caption{Swap count per scenario for H1 and H2 (Batch~2; moving average over 7 scenarios).
        }
        \label{fig:mpc2swaps}
\end{figure}

Figure~\ref{fig:mpc2swaps} compares H1 and H2: H2 requires fewer swaps than in Batch~1, consistent with the simpler conflict structure, while the swap count of H1 remains essentially unchanged. The advantage of H1 over H2 is maintained across all instances.

Finally, Table~\ref{tab:batch_summary} summarizes the key metrics across both batches.
Wider inter-arrival gaps consistently reduce scheduling complexity: the mean
optimality gap falls from $18.8\%$ to $3.9\%$, and the swap count of H2 decreases by
roughly $20\%$.
Across both batches, H1 strictly dominates H2 in computational cost while
delivering equal or better solution quality, confirming the persistent swap
memory~$\mathcal{T}$ as an effective and reliable mechanism for reducing
redundant NLP solves in the dynamic setting.

\begin{table}[!tbh]
    \centering
    \caption{Summary of statistical evaluation across both batches (6 vehicles,
    100 scenarios each).}
    \label{tab:batch_summary}
    \begin{tabularx}{\columnwidth}{>{\hsize=1.2\hsize}X
                                   >{\centering\arraybackslash\hsize=0.9\hsize}X
                                   >{\centering\arraybackslash\hsize=0.9\hsize}X}
        \toprule
        Metric & Batch~1 & Batch~2 \\
                        & $[0.2,0.4]$\,[s]   & $[0.3,0.5]$\,[s]   \\
        \midrule
        Mean OGR (vs FIFO)        & $85.5\%$ & $87.0\%$ \\
        Mean OG (vs optimal)      & $18.8\%$ & $3.9\%$  \\
        H1 swaps (range)          & $7$--$10$   & $5$--$9$    \\
        H2 swaps (range)          & $15$--$19$   & $11$--$16$    \\
        \bottomrule
    \end{tabularx}
\end{table}

\subsection{Continuous arrival evaluation}
\label{sec:zone_release}

To assess the proposed heuristic beyond the fixed-size instances considered in the
previous subsections, a $300$\,s closed-loop simulation with a continuous stream of
arrivals is carried out. The vehicle dynamics, cost structure, and class weights are those
of the preceding experiments, and the proposed scheme is run with H1 ($\alpha = 1$).
Two traffic densities are considered, a \emph{sparse} and a \emph{dense} regime,
distinguished by the inter-arrival gap reported in Table~\ref{tab:zone_release}.

The proposed heuristic and FIFO are evaluated in two independent closed loops driven
by the same arrival sequence. In both loops, \eqref{eq:coordination_ocp} is re-solved
whenever a vehicle enters the coordination set or leaves the intersection. FIFO
retains the arrival order throughout, whereas the proposed heuristic appends each
new vehicle to the current order and applies the multiplier-guided swaps. Each
strategy therefore evolves its own trajectories, and performance is measured by
the closed-loop cost actually incurred by the vehicles, accumulated over the
simulation and normalized per coordinated vehicle.

\begin{figure}[!t]
  \centering
  \includegraphics[width=\linewidth]{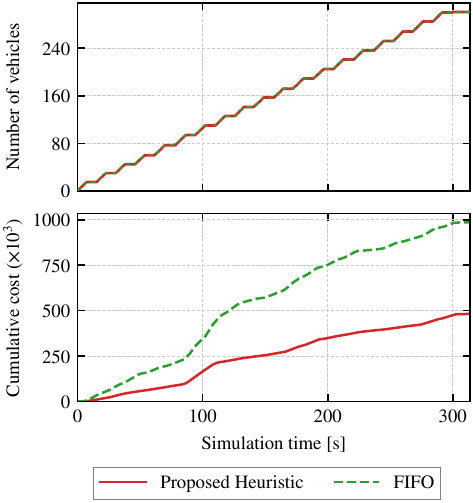}
  \caption{Sparse regime ($301$ vehicles, inter-arrival gaps in $[0.4,0.6]$\,[s]).
  Top: cumulative number of vehicles entering the coordination set.
  Bottom: cumulative closed-loop cost of the proposed heuristic and FIFO.}
  \label{fig:zone_sparse}
\end{figure}

\begin{figure}[!t]
  \centering
  \includegraphics[width=\linewidth]{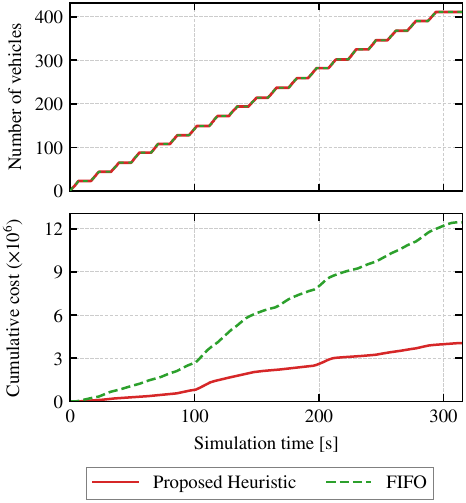}
  \caption{Dense regime ($411$ vehicles, inter-arrival gaps in $[0.2,0.4]$\,[s]).
  Top: cumulative number of vehicles entering the coordination set.
  Bottom: cumulative closed-loop cost of the proposed heuristic and FIFO.}
  \label{fig:zone_dense}
\end{figure}

\begin{table}[!tbh]
\centering
\caption{Proposed heuristic vs FIFO across two traffic densities}
\label{tab:zone_release}
\begin{tabularx}{\columnwidth}{>{\hsize=1.3\hsize}X
                               >{\centering\arraybackslash\hsize=0.85\hsize}X
                               >{\centering\arraybackslash\hsize=0.85\hsize}X}
\toprule
Metric & Sparse & Dense \\
\midrule
Arrival gap                     & $[0.4,0.6]$\,[s] & $[0.2,0.4]$\,[s] \\
Coordinated vehicles            & $301$       & $411$       \\
Mean cost (FIFO)                & $3.27\times10^{3}$ & $3.03\times10^{4}$ \\
Mean cost (proposed)            & $1.60\times10^{3}$ & $9.87\times10^{3}$ \\
Mean cost reduction vs FIFO     & $51.1\%$    & $67.4\%$    \\
Accepted swaps                  & $332$       & $963$       \\
Rejected swaps                  & $301$       & $410$       \\
Swap acceptance rate            & $52.4\%$    & $70.1\%$    \\
\bottomrule
\end{tabularx}
\end{table}

Table~\ref{tab:zone_release} summarizes the results, and
Figures~\ref{fig:zone_sparse} and~\ref{fig:zone_dense} show their evolution in time. All $301$ and $411$ vehicles are coordinated and cross the intersection under both
strategies, without any violation of the precedence constraint. The proposed
heuristic reduces the mean closed-loop cost relative to FIFO by $51.1\%$ in the
sparse regime and by $67.4\%$ in the dense regime. The cumulative cost curves
diverge from the first arrivals onward and the gap widens steadily over the
horizon, showing that the improvement is sustained along the stream rather than
concentrated in isolated events. The reduction grows with density: the mean FIFO
cost increases by roughly an order of magnitude from the sparse to the dense regime,
whereas the cost of the proposed heuristic grows considerably less.

The swap statistics show that the candidate selection remains effective at scale.
The acceptance rate is $52.4\%$ in the sparse regime and $70.1\%$ in the dense one.
The higher rate under dense traffic is consistent with larger coordination sets
offering more improving reorderings among the multiplier-ranked candidates. Together, these results extend the conclusions drawn from the fixed-size instances
to a sustained stream of several hundred vehicles, where enumeration-based baselines
are intractable.

\section{Conclusion}

We have presented a receding-horizon NLP framework for the online coordination of autonomous vehicles at unsignalized intersections under random arrivals. Building on the static Lagrange-multiplier-based reordering heuristic of~\cite{shahi2026lagrange}, we extended the approach to the dynamic setting by embedding it inside an arrival-triggered MPC loop. The framework relies on three design choices: new vehicles are appended at the tail of the current crossing order and warm-started from the previous NLP solution, the swap memory $\mathcal{T}$ prevents redundant NLP solves across replanning instants, and candidate swaps are ranked by the KKT multipliers associated with the precedence constraints, so that computational effort focuses on orderings most likely to yield improvement. Extensive numerical results on several dynamic traffic scenarios of interest have shown how the proposed technique avoids combinatorial enumeration and relies only on repeated solutions of a continuous NLP, whose size scales polynomially with the number of vehicles, while bridging the gap between FIFO and the brute-force global optimum obtained via enumeration. 

A natural extension of this work concerns real-time execution. The present scheme assumes that each replanning terminates within the prescribed time $t_\mathrm{alg}$, whereas every candidate swap still requires a full NLP solve, whose time grows with the number of vehicles. This limitation can be addressed by exploiting the separable structure of~\eqref{eq:coordination_ocp} through a bi-level decomposition, in which each vehicle solves its own optimal control problem and a coordinator optimizes the crossing times. Parametric sensitivities of the vehicle problems can then be used to evaluate candidate swaps through a quadratic approximation instead of repeated NLP solves, so that the computation time can be accounted for explicitly in closed loop and the multiplier-guided reordering can be executed within the sampling time.

\bibliographystyle{IEEEtran}
\bibliography{references}	

@ARTICLE{namazi2019intelligent,
  author={Namazi, Elnaz and Li, Jingyue and Lu, Chaoru},
  journal={IEEE Access}, 
  title={Intelligent Intersection Management Systems Considering Autonomous Vehicles: A Systematic Literature Review}, 
  year={2019},
  volume={7},
  number={},
  pages={91946-91965},
  doi={10.1109/ACCESS.2019.2927412}}

@ARTICLE{wu2022intersection,
  author={Wu, Jiaming and Qu, Xiaobo},
  journal={Journal of Intelligent and Connected Vehicles}, 
  title={Intersection control with connected and automated vehicles: A review}, 
  year={2022},
  volume={5},
  number={3},
  pages={260-269},
  doi={10.1108/JICV-06-2022-0023}}

@ARTICLE{rios2016survey,
  author={Rios-Torres, Jackeline and Malikopoulos, Andreas A.},
  journal={IEEE Transactions on Intelligent Transportation Systems}, 
  title={A Survey on the Coordination of Connected and Automated Vehicles at Intersections and Merging at Highway On-Ramps}, 
  year={2017},
  volume={18},
  number={5},
  pages={1066-1077},
  doi={10.1109/TITS.2016.2600504}}

@ARTICLE{al2022signalized,
  author={Al-Turki, Mohammed and Ratrout, Nedal T. and Rahman, Syed Masiur and Assi, Khaled J.},
  journal={IEEE Access}, 
  title={Signalized Intersection Control in Mixed Autonomous and Regular Vehicles Traffic Environment—A Critical Review Focusing on Future Control}, 
  year={2022},
  volume={10},
  number={},
  pages={16942-16951},
  doi={10.1109/ACCESS.2022.3148706}}

@ARTICLE{fayazi2018mixed,
  author={Fayazi, Seyed Alireza and Vahidi, Ardalan},
  journal={IEEE Transactions on Intelligent Vehicles}, 
  title={Mixed-Integer Linear Programming for Optimal Scheduling of Autonomous Vehicle Intersection Crossing}, 
  year={2018},
  volume={3},
  number={3},
  pages={287-299},
  doi={10.1109/TIV.2018.2843163}}

@article{muller2016intersection,
title = {Intersection control for automated vehicles with {MILP}},
journal = {IFAC-PapersOnLine},
volume = {49},
number = {3},
pages = {37-42},
year = {2016},
issn = {2405-8963},
doi = {https://doi.org/10.1016/j.ifacol.2016.07.007},
  author = {M{\"u}ller, Eduardo Rauh and Carlson, Rodrigo Castelan and Kraus Jr., Werner},
  note = {14th IFAC Symposium on Control in Transportation Systems (CTS 2016)},
}

@INPROCEEDINGS{fayazi2017optimal,
  author={Fayazi, S. Alireza and Vahidi, Ardalan and Luckow, Andre},
  booktitle={2017 American Control Conference (ACC)}, 
  title={Optimal scheduling of autonomous vehicle arrivals at intelligent intersections via {MILP}}, 
  year={2017},
  volume={},
  number={},
  pages={4920-4925},
  doi={10.23919/ACC.2017.7963717}}

@article{yu2018integrated,
title = {Integrated optimization of traffic signals and vehicle trajectories at isolated urban intersections},
journal = {Transportation Research Part B: Methodological},
volume = {112},
pages = {89-112},
year = {2018},
issn = {0191-2615},
doi = {https://doi.org/10.1016/j.trb.2018.04.007},
author = {Chunhui Yu and Yiheng Feng and Henry X. Liu and Wanjing Ma and Xiaoguang Yang},

}

@ARTICLE{fabiani2019mixed,
  author={Fabiani, Filippo and Grammatico, Sergio},
  journal={IEEE Transactions on Intelligent Transportation Systems}, 
  title={Multi-Vehicle Automated Driving as a Generalized Mixed-Integer Potential Game}, 
  year={2020},
  volume={21},
  number={3},
  pages={1064-1073},
  doi={10.1109/TITS.2019.2901505}}

@INPROCEEDINGS{qian2014priority,
  author={Qian, Xiangjun and Gregoire, Jean and Moutarde, Fabien and De La Fortelle, Arnaud},
  booktitle={17th International IEEE Conference on Intelligent Transportation Systems (ITSC)}, 
  title={Priority-based coordination of autonomous and legacy vehicles at intersection}, 
  year={2014},
  volume={},
  number={},
  pages={1166-1171},
  doi={10.1109/ITSC.2014.6957845}}

@ARTICLE{zhang2015state,
  author={Zhang, Kailong and Zhang, Dafang and de La Fortelle, Arnaud and Wu, Xiao and Grégoire, Jean},
  journal={IEEE Transactions on Intelligent Transportation Systems}, 
  title={State-Driven Priority Scheduling Mechanisms for Driverless Vehicles Approaching Intersections}, 
  year={2015},
  volume={16},
  number={5},
  pages={2487-2500},
  doi={10.1109/TITS.2015.2411619}}

@article{dresner2008multiagent,
author = {Dresner, Kurt and Stone, Peter},
year = {2008},
pages = {591-656},
title = {A Multiagent Approach to Autonomous Intersection Management},
volume = {31},
journal = {Journal of Artificial Intelligence Research},
doi = {10.1613/jair.2502}
}

@article{levin2017conflict,
title = {Conflict-point formulation of intersection control for autonomous vehicles},
journal = {Transportation Research Part C: Emerging Technologies},
author = {Michael W. Levin and David Rey},
volume = {85},
pages = {528-547},
year = {2017},
issn = {0968-090X},
doi = {https://doi.org/10.1016/j.trc.2017.09.025},
}

@ARTICLE{levin2016optimizing,
  author={Levin, Michael W. and Fritz, Hagen and Boyles, Stephen D.},
  journal={IEEE Transactions on Intelligent Transportation Systems}, 
  title={On Optimizing Reservation-Based Intersection Controls}, 
  year={2017},
  volume={18},
  number={3},
  pages={505-515},
  doi={10.1109/TITS.2016.2574948}}

@inproceedings{Hult2016PrimalDecomp,
author={Hult, Robert and Zanon, Mario and Gros, Sébastien and Falcone, Paolo},
  booktitle={2016 IEEE 55th Conference on Decision and Control (CDC)}, 
  title={Primal decomposition of the optimal coordination of vehicles at traffic intersections}, 
  year={2016},
  volume={},
  number={},
  pages={2567-2573},
  doi={10.1109/CDC.2016.7798649}}

@article{Zanon2017Asynchronous,
title = {An Asynchronous Algorithm for Optimal Vehicle Coordination at Traffic Intersections},
journal = {IFAC-PapersOnLine},
volume = {50},
number = {1},
pages = {12008-12014},
year = {2017},
note = {20th IFAC World Congress},
issn = {2405-8963},
doi = {https://doi.org/10.1016/j.ifacol.2017.08.2124},
author = {Mario Zanon and Sébastien Gros and Henk Wymeersch and Paolo Falcone}}

@ARTICLE{hult2019optimal,
  author={Hult, Robert and Zanon, Mario and Gros, Sebastien and Falcone, Paolo},
  journal={IEEE Transactions on Control Systems Technology}, 
  title={Optimal Coordination of Automated Vehicles at Intersections: Theory and Experiments}, 
  year={2019},
  volume={27},
  number={6},
  pages={2510-2525},
  doi={10.1109/TCST.2018.2871397}}

@INPROCEEDINGS{Hult2018ECC,
  author={Hult, Robert and Zanon, Mario and Gros, Sébastien and Falcone, Paolo},
  booktitle={2018 European Control Conference (ECC)}, 
  title={Energy-Optimal Coordination of Autonomous Vehicles at Intersections}, 
  year={2018},
  volume={},
  number={},
  pages={602-607},
  doi={10.23919/ECC.2018.8550367}}

@article{Hult2020Coordination,
author = {Robert Hult and Mario Zanon and Sébastien Gros and Henk Wymeersch and Paolo Falcone},
title = {Optimisation-based coordination of connected, automated vehicles at intersections},
journal = {Vehicle System Dynamics},
volume = {58},
number = {5},
pages = {726--747},
year = {2020},
publisher = {Taylor \& Francis},
doi = {10.1080/00423114.2020.1755446},
}

@article{Hult2020Experimental,
author = {Hult, Robert and Zanon, Mario and Frison, Gianluca and Gros, Sébastien and Falcone, Paolo},
title = {Experimental validation of a semi-distributed sequential quadratic programming method for optimal coordination of automated vehicles at intersections},
journal = {Optimal Control Applications and Methods},
volume = {41},
number = {4},
pages = {1068-1096},
doi = {https://doi.org/10.1002/oca.2592},
year = {2020}
}

@ARTICLE{Hult2022Intersection,
  author={Hult, Robert and Zanon, Mario and Gros, Sébastien and Falcone, Paolo},
  journal={IEEE Transactions on Control Systems Technology}, 
  title={A Semidistributed Interior Point Algorithm for Optimal Coordination of Automated Vehicles at Intersections}, 
  year={2022},
  volume={30},
  number={5},
  pages={1977-1989},
  doi={10.1109/TCST.2021.3132835}}

@INPROCEEDINGS{Hult2018CDC,
  author={Hult, Robert and Zanon, Mario and Gros, Sebastien and Falcone, Paolo},
  booktitle={2018 IEEE Conference on Decision and Control (CDC)}, 
  title={An {MIQP}-based heuristic for Optimal Coordination of Vehicles at Intersections}, 
  year={2018},
  volume={},
  number={},
  pages={2783-2790},
  doi={10.1109/CDC.2018.8618945}}

@INPROCEEDINGS{Murgovski2015Convex,
  author={Murgovski, Nikolce and de Campos, Gabriel Rodrigues and Sjöberg, Jonas},
  booktitle={2015 54th IEEE Conference on Decision and Control (CDC)}, 
  title={Convex modeling of conflict resolution at traffic intersections}, 
  year={2015},
  volume={},
  number={},
  pages={4708-4713},
  doi={10.1109/CDC.2015.7402953}}

@INPROCEEDINGS{deCampos2014Cooperative,
  author={de Campos, Gabriel Rodrigues and Falcone, Paolo and Wymeersch, Henk and Hult, Robert and Sjöberg, Jonas},
  booktitle={53rd IEEE Conference on Decision and Control}, 
  title={Cooperative receding horizon conflict resolution at traffic intersections}, 
  year={2014},
  volume={},
  number={},
  pages={2932-2937},
  doi={10.1109/CDC.2014.7039840}}

@article{KimKumar2014,
author={Kim, Kyoung-Dae and Kumar, P. R.},
  journal={IEEE Transactions on Automatic Control}, 
  title={An {MPC}-Based Approach to Provable System-Wide Safety and Liveness of Autonomous Ground Traffic}, 
  year={2014},
  volume={59},
  number={12},
  pages={3341-3356},
  doi={10.1109/TAC.2014.2351911}}

@article{Katriniok2017Distributed,
title = {Distributed Model Predictive Control for Intersection Automation Using a Parallelized Optimization Approach},
journal = {IFAC-PapersOnLine},
volume = {50},
number = {1},
pages = {5940-5946},
year = {2017},
note = {20th IFAC World Congress},
issn = {2405-8963},
doi = {https://doi.org/10.1016/j.ifacol.2017.08.1492},
author = {Alexander Katriniok and Peter Kleibaum and Martina Joševski},
}

@article{zhang2016optimal,
title = {Decentralized optimal control of Connected Automated Vehicles at signal-free intersections including comfort-constrained turns and safety guarantees},
journal = {Automatica},
volume = {109},
pages = {108563},
year = {2019},
issn = {0005-1098},
doi = {https://doi.org/10.1016/j.automatica.2019.108563},
author = {Yue Zhang and Christos G. Cassandras},
}

@article{malikopoulos2018decentralized,
title = {A decentralized energy-optimal control framework for connected automated vehicles at signal-free intersections},
journal = {Automatica},
volume = {93},
pages = {244-256},
year = {2018},
issn = {0005-1098},
doi = {https://doi.org/10.1016/j.automatica.2018.03.056},
author = {Andreas A. Malikopoulos and Christos G. Cassandras and Yue J. Zhang},
}

@INPROCEEDINGS{rios2017online,
  author={Rios-Torres, Jackeline and Malikopoulos, Andreas and Pisu, Pierluigi},
  booktitle={2015 IEEE 18th International Conference on Intelligent Transportation Systems}, 
  title={Online Optimal Control of Connected Vehicles for Efficient Traffic Flow at Merging Roads}, 
  year={2015},
  volume={},
  number={},
  pages={2432-2437},
  doi={10.1109/ITSC.2015.392}}

@ARTICLE{chalaki2021optimal,
  author={Chalaki, Behdad and Malikopoulos, Andreas A.},
  journal={IEEE Transactions on Control Systems Technology}, 
  title={Optimal Control of Connected and Automated Vehicles at Multiple Adjacent Intersections}, 
  year={2022},
  volume={30},
  number={3},
  pages={972-984},
  doi={10.1109/TCST.2021.3082306}}

@ARTICLE{Faris2024TCST,
  author={Faris, Muhammad and Zanon, Mario and Falcone, Paolo},
  journal={IEEE Transactions on Control Systems Technology}, 
  title={An Optimization-Based Dynamic Reordering Heuristic for Coordination of Vehicles in Mixed Traffic Intersections}, 
  year={2025},
  volume={33},
  number={4},
  pages={1387-1402},
  doi={10.1109/TCST.2024.3508542}}

@INPROCEEDINGS{Scheffe2022DynamicPriority,
  author={Scheffe, Patrick and Dorndorf, Georg and Alrifaee, Bassam},
  booktitle={2022 IEEE 25th International Conference on Intelligent Transportation Systems (ITSC)}, 
  title={Increasing Feasibility with Dynamic Priority Assignment in Distributed Trajectory Planning for Road Vehicles}, 
  year={2022},
  volume={},
  number={},
  pages={3873-3879},
  doi={10.1109/ITSC55140.2022.9922028}}

@misc{shahi2026lagrange,
  author       = {Shahi, Saugat and Fabiani, Filippo and Zanon, Mario},
  title        = {A Heuristic for Scheduling Autonomous Vehicles at
Intersections},
  year         = {2026},
  howpublished = {arXiv:2610.37275 [eess.SY]},
  url          = {https://arxiv.org/abs/2610.37275}
}

@ARTICLE{funke2017,
  author={Funke, Joseph and Brown, Matthew and Erlien, Stephen M. and Gerdes, J. Christian},
  journal={IEEE Transactions on Control Systems Technology}, 
  title={Collision Avoidance and Stabilization for Autonomous Vehicles in Emergency Scenarios}, 
  year={2017},
  volume={25},
  number={4},
  pages={1204-1216},
  doi={10.1109/TCST.2016.2599783}}

@ARTICLE{zhou2019distributed,
  author={Zheng, Yang and Li, Shengbo Eben and Li, Keqiang and Borrelli, Francesco and Hedrick, J. Karl},
  journal={IEEE Transactions on Control Systems Technology}, 
  title={Distributed Model Predictive Control for Heterogeneous Vehicle Platoons Under Unidirectional Topologies}, 
  year={2017},
  volume={25},
  number={3},
  pages={899-910},
  doi={10.1109/TCST.2016.2594588}}

@article{diehl2005real,
author = {Diehl, Moritz and Bock, Hans Georg and Schl{\"o}der, Johannes P.},
year = {2005},
number = {5},
pages = {1714-1736},
title = {A Real-Time Iteration Scheme for Nonlinear Optimization in Optimal Feedback Control},
volume = {43},
journal = {SIAM Journal on Control and Optimization},
doi = {10.1137/S0363012902400713}
}

@book{nocedal2006numerical,
  title={Numerical Optimization},
  author={Nocedal, Jorge and Wright, Stephen J.},
  edition={2nd},
  year={2006},
  publisher={Springer},
  series={Springer Series in Operations Research and Financial Engineering},
  address={New York, NY, USA},
  doi={10.1007/978-0-387-40065-5}
}

@article{Andersson2019CasADi,
author = {Andersson, Joel A. E. and Gillis, Joris and Horn, Greg and Rawlings, James B. and Diehl, Moritz},
year = {2019},
number = {1},
pages = {1-36},
title = {{CasADi}: a software framework for nonlinear optimization and optimal control},
volume = {11},
journal = {Mathematical Programming Computation},
doi = {10.1007/s12532-018-0139-4}
}

@article{wachter2006implementation,
author = {Wächter, Andreas and Biegler, Lorenz T.},
year = {2006},
number = {1},
pages = {25-57},
title = {On the implementation of an interior-point filter line-search algorithm for large-scale nonlinear programming},
volume = {106},
journal = {Mathematical Programming},
doi = {10.1007/s10107-004-0559-y}
}

@IEEEtranBSTCTL{IEEEexample:BSTcontrol,
  CTLdash_repeated_names = "no"
}
\end{document}